\documentclass[useAMS,usenatbib]{mn2e}
\usepackage{times}
\usepackage{graphicx}
\usepackage{subfigure}
\usepackage{psfrag}
\usepackage{amsmath}
\usepackage{amssymb}

\def\reff@jnl#1{{\rm#1\/}}
\def\aj{\reff@jnl{AJ}}                 
\def\araa{\reff@jnl{ARA\&A}}           
\def\apj{\reff@jnl{ApJ}}               
\def\apjl{\reff@jnl{ApJ}}              
\def\apjs{\reff@jnl{ApJS}}             
\def\ao{\reff@jnl{Appl.Optics}}        
\def\apss{\reff@jnl{Ap\&SS}}           
\def\aap{\reff@jnl{A\&A}}              
\def\aapr{\reff@jnl{A\&A~Rev.}}        
\def\aaps{\reff@jnl{A\&AS}}            
\def\azh{\reff@jnl{AZh}}               
\def\baas{\reff@jnl{BAAS}}             
\def\jrasc{\reff@jnl{JRASC}}           
\def\memras{\reff@jnl{MmRAS}}          
\def\mnras{\reff@jnl{MNRAS}}           
\def\pra{\reff@jnl{Phys.Rev.A}}        
\def\prb{\reff@jnl{Phys.Rev.B}}        
\def\prc{\reff@jnl{Phys.Rev.C}}        
\def\prd{\reff@jnl{Phys.Rev.D}}        
\def\prl{\reff@jnl{Phys.Rev.Lett}}     
\def\pasp{\reff@jnl{PASP}}             
\def\pasj{\reff@jnl{PASJ}}             
\def\qjras{\reff@jnl{QJRAS}}           
\def\skytel{\reff@jnl{S\&T}}           
\def\solphys{\reff@jnl{Solar~Phys.}}   
\def\sovast{\reff@jnl{Soviet~Ast.}}    
\def\ssr{\reff@jnl{Space~Sci.Rev.}}    
\def\zap{\reff@jnl{ZAp}}               
\def\nat{\reff@jnl{Nature}}            

\title[Cluster detection in weak lensing surveys]{Cluster detection in weak lensing surveys} 
\author[F. Feroz, P.J. Marshall and M.P. Hobson] 
{F. Feroz$^{1}$\thanks{E-mail: f.feroz@mrao.cam.ac.uk}, P.J. Marshall$^{2}$ and M.P.~Hobson$^{1}$\\ 
$^{1}$Astrophysics Group, Cavendish Laboratory, JJ Thomson Avenue, Cambridge CB3 0HE, UK\\
$^{2}$UC Santa Barbara, Santa Barbara CA 93106, USA }

\date{Accepted ---. Received ---; in original form \today}
\pagerange{\pageref{firstpage}--\pageref{lastpage}}
\pubyear{2008}

\begin{document}
\label{firstpage}
\maketitle

\begin{abstract}

We present an efficient and robust approach for extracting clusters of galaxies from weak lensing survey data and
measuring their properties. We use simple, physically-motivated cluster models appropriate for such sparse, noisy
data, and incorporate our knowledge of the cluster mass function to optimise the detection of low-mass objects.
Despite the method's non-linear nature, we are able to search at a rate of approximately half a square degree per
hour on a single processor, making this technique a viable candidate for future wide-field surveys. We quantify,
for two simulated data-sets, the accuracy of recovered cluster parameters, and discuss the completeness and
purity of our shear-selected cluster catalogues.

\end{abstract}


\section{Introduction}\label{sec:intro}

Clusters of galaxies are the most massive gravitationally bound objects in the universe and, as such, are
critical tracers of the formation of large-scale structure. The number count of clusters as a function of their
mass and redshift has been predicted both analytically (see e.g. \citealt{press74,sheth01}) and from large scale
numerical simulations (see e.g. \citealt{jenkins01,evrard02}), and are particularly sensitive to the cosmological
parameters $\sigma_8$ and $\Omega_{\rm m}$ (see e.g. \citealt{battye03}). The size and formation history of massive
clusters is such that the ratio of gas mass to total mass is expected to be representative of the universal ratio
$\Omega_{\rm b}/\Omega_{\rm m}$, once the relatively small amount of baryonic matter in the cluster galaxies is
taken into account (see e.g. \citealt{white93}).

The study of cosmic shear has rapidly progressed with the availability of high quality wide-field lensing data. 
Large dedicated surveys with ground-based telescopes have been employed to reconstruct the mass distribution of
the universe and constrain cosmological parameters (see e.g. \citealt{massey07, massey05, hoekstra06}).  Weak
lensing also allows one to detect galaxy clusters without making any assumptions about the baryon fraction,
richness, morphology or dynamical state of the cluster, and so weak lensing cluster modelling would allow one to
test these assumptions by observing the cluster with optical, X-ray or Sunyaev-Zel'dovich (SZ) methods.

Despite the advances in data quality, weak lensing data remains very sparse and noisy. Hence, obtaining the shear
signal from the shapes of the background galaxies is a very challenging task. Several grid-based methods have
been devised to reconstruct the projected mass distribution from the observed shear field (see
e.g. \citealt{kaiser93, squires96, bridle99, starck06}). In these ``non-parametric'' methods, the model assumed is
a grid of pixels whose values comprise the model parameters. \citet{marshall02} showed that such a large number of
parameters is often not justified by the data quality -- failure to recognise this can result in over-fitting
and over-interpretation of the data.

An alternative method for mass reconstruction is to work with simply-parameterised physically-motivated models
for the underlying mass distribution \citep{marshall03}.  By fitting simple template cluster models to the
observed data set, we can draw inferences about the cluster parameters directly. This involves calculating the
probability distribution of these parameters, and also (perhaps) those of the background cosmology; we can also
compare different cluster models, enhancing our astrophysical understanding of these systems. This is most
conveniently done through a Bayesian inference.

In \citet{marshall03} and \citet{marshall06} a Bayesian approach was presented for such an analysis of weak
lensing data from pointed observations towards known clusters. This used a highly effective, but computationally
intensive, Markov Chain Monte Carlo (MCMC) sampler to explore the high-dimensional parameter space, and employed
the thermodynamic integration technique to calculate the Bayesian evidence. In this paper, we extend this work by
utilizing the recently developed `multimodal nested sampling' ({\sc MultiNest}) technique
\citep{feroz08,feroz08b}, which is found to be $\sim 200$ times more efficient than traditional MCMC methods and
thus enables one to search for multiple clusters in wide-field weak lensing data. {\sc MultiNest} enables one to
simultaneously detect clusters from the weak lensing data and perform the parameter estimation for the individual
cluster model parameters. Furthermore, following \citet{hobson03}, we also quantify our cluster detection through
the application of Bayesian model selection using the Bayesian evidence value for each detected cluster, which
can be easily calculated using the {\sc MultiNest} technique.

The outline of this paper is as follows. In Section \ref{sec:method} we describe our methodology for detecting
and characterising clusters in weak lensing survey data. In Section \ref{sec:simpledemo} we apply our method to a
simple simulated data-set, before moving on to describe realistic cluster survey simulations and the results of
our cluster extraction algorithm on these simulations in Section \ref{sec:whitedemo}. Finally we present our
conclusions in Section \ref{sec:conclusions}.

\section{Methodology}\label{sec:method}

\subsection{Bayesian inference}\label{sec:method:bayesian}

Our cluster detection methodology is built upon the principles of
Bayesian inference, and so we begin by giving a brief summary of this
framework. Bayesian inference methods provide a consistent approach to
the estimation of a set of parameters $\mathbf{\Theta}$ in a model (or
hypothesis) $H$ for the data $\mathbf{D}$. Bayes' theorem states that
\begin{equation} \Pr(\mathbf{\Theta}|\mathbf{D}, H) =
\frac{\Pr(\mathbf{D}|\,\mathbf{\Theta},H)
\Pr(\mathbf{\Theta}|H)}{\Pr(\mathbf{D}|H)},
\end{equation}
where $\Pr(\mathbf{\Theta}|\mathbf{D}, H) \equiv P(\mathbf{\Theta})$
is the posterior probability distribution of the parameters,
$\Pr(\mathbf{D}|\mathbf{\Theta}, H) \equiv \mathcal{L}(\mathbf{\Theta})$ is the
likelihood, $\Pr(\mathbf{\Theta}|H) \equiv \pi(\mathbf{\Theta})$ is
the prior, and $\Pr(\mathbf{D}|H) \equiv \mathcal{Z}$ is the Bayesian
evidence.

In parameter estimation, the normalising evidence factor is usually
ignored, since it is independent of the parameters $\mathbf{\Theta}$,
and inferences are obtained by taking samples from the (unnormalised)
posterior using standard MCMC sampling methods, where at equilibrium
the chain contains a set of samples from the parameter space
distributed according to the posterior. This posterior constitutes the
complete Bayesian inference of the parameter values, and can be
marginalised over each parameter to obtain individual parameter
constraints. 

In contrast to parameter estimation problems, for model selection the
evidence takes the central role and is simply the factor required to
normalize the posterior over $\mathbf{\Theta}$:
\begin{equation}
\mathcal{Z} =
\int{\mathcal{L}(\mathbf{\Theta})\pi(\mathbf{\Theta})}d^D\mathbf{\Theta},
\label{eq:3}
\end{equation} 
where $D$ is the dimensionality of the parameter space. As the average
of the likelihood over the prior, the evidence is larger for a model
if more of its parameter space is likely and smaller for a model with
large areas in its parameter space having low likelihood values, even
if the likelihood function is very highly peaked. Thus, the evidence
automatically implements Occam's razor: a simpler theory with compact
parameter space will have a larger evidence than a more complicated
one, unless the latter is significantly better at explaining the data.
The question of model selection between two models $H_{0}$ and $H_{1}$
can then be decided by comparing their respective posterior
probabilities given the observed data set $\mathbf{D}$, as follows
\begin{equation}
R=\frac{\Pr(H_{1}|\mathbf{D})}{\Pr(H_{0}|\mathbf{D})}
=\frac{\Pr(\mathbf{D}|H_{1})\Pr(H_{1})}{\Pr(\mathbf{D}|
H_{0})\Pr(H_{0})}
=\frac{\mathcal{Z}_1}{\mathcal{Z}_0}\frac{\Pr(H_{1})}{\Pr(H_{0})},
\label{eq:3.1}
\end{equation}
where $\Pr(H_{1})/\Pr(H_{0})$ is the a priori probability ratio for
the two models, which can often be set to unity but occasionally
requires further consideration.

Evaluation of the multidimensional integral in Eq.~\ref{eq:3} is a
challenging numerical task. Standard techniques like thermodynamic
integration are extremely computationally expensive which makes
evidence evaluation at least an order of magnitude more costly than
parameter estimation. Some fast approximate methods have been used for
evidence evaluation, such as treating the posterior as a multivariate
Gaussian centred at its peak (see e.g. \citealt{hobson03}), but this
approximation is clearly a poor one for multimodal posteriors (except
perhaps if one performs a separate Gaussian approximation at each
mode). The Savage-Dickey density ratio has also been proposed (see
e.g. \citealt{trotta05}) as an exact, and potentially faster, means of
evaluating evidences, but is restricted to the special case of nested
hypotheses and a separable prior on the model parameters. Various
alternative information criteria for astrophysical model selection are
discussed by \citet{liddle07}, but the evidence remains the preferred
method.

The nested sampling approach, introduced by \citet{skilling04}, is a
Monte Carlo method targeted at the efficient calculation of the
evidence, but also produces posterior inferences as a
by-product. \citet{feroz08} and \citet{feroz08b} built on this nested
sampling framework and have recently introduced the {\sc MultiNest}
algorithm which is very efficient in sampling from posteriors that may
contain multiple modes and/or large (curving) degeneracies and also
calculates the evidence. This technique has greatly reduced the
computational cost of Bayesian parameter estimation and
model selection, and is employed in this paper.

\subsection{Weak lensing likelihood}\label{sec:method:lensing}

Our approach to detecting multiple clusters in weak lensing data
follows the generic object detection strategy advocated by
\citet{hobson03} and refined by \citet{feroz08}. They show that the
straightforward approach of using a single-object model for the data
is both computationally far less demanding than adopting a
multiple-object model and reliable, provided that the objects of
interest are spatially well-separated.  It is important to understand
that adopting a single-object model does {\em not} restrict one to
detecting just a single cluster in the weak lensing data.  Rather, by
modelling the data in this way, one expects the posterior distribution
to possess local maxima in the parameter space $\mathbf{\Theta}$ of
the single-cluster model, some of which will correspond to real
clusters present in the data and some that occur because the pattern
of ellipticities in the background galaxies `conspire' to give the
impression that a cluster might be present.  The process of object
detection and characterisation thus reduces to locating the local
maxima of the posterior distribution in the parameter space
$\mathbf{\Theta}$ and deciding which of these local maxima correspond
to a real cluster.

A model cluster density profile can be determined from numerical
$N$-body simulations of large-scale structure formation in a
$\Lambda$CDM universe.  In particular, assuming spherical symmetry,
the NFW profile (\citealt{navarro97}) provides a good fit to the
simulations and is given by
\begin{equation}
\rho(r)=\frac{\rho_{\rm s}}{(r/r_{\rm s})(1+r/r_{\rm s})^2},
\label{eq:NFW}
\end{equation}
where $r_{\rm s}$ and $\rho_{\rm s}$ are the radius and density at
which the logarithmic slope breaks from $-1$ to $-3$. The mass
$M_{200}$ contained within the radius $r_{200}$ at which the density
is $200$ times the cosmological critical density $\rho_{\rm crit}$ at
the redshift of the cluster can be calculated as (\citealt{evrard02,
  allen03}):
\begin{eqnarray}
\frac{M_{200}}{(4/3)\pi r_{200}^3}&=&200\rho_{\rm crit}, \nonumber \\
&=&4\pi \rho_{\rm s} r_{\rm s}^3\left[\log(1+c)-\frac{c}{1+c}\right],
\label{eq:rho200}
\end{eqnarray}
where $c=r_{200}/r_{\rm s}$ is a measure of the halo concentration.
Thus, we take as our cluster parameters $\mathbf{\Theta} = (x_{\rm
  c},y_{\rm c},M_{200},c,z)$, where $x_{\rm c}$ and $y_{\rm c}$ are
the spatial coordinates at which the cluster is centred, and $z$ is
its redshift.

A cluster mass distribution is investigated using weak gravitational
lensing through the relationship (see e.g. \citealt{schramm95}):
\begin{equation}
\langle\epsilon(\bmath{x})\rangle=g(\bmath{x}),
\end{equation}
that is, at any point $\bmath{x}$ on the sky, the local average of the
complex ellipticities $\epsilon=\epsilon_1+i\epsilon_2$ of a
collection of background galaxy images is an unbiased estimator of the
local complex reduced shear field, $g=g_1+ig_2$, due to the cluster.
Adopting the thin-lens approximation, for a projected mass
distribution $\Sigma(\bmath{x})$ in the lens, the reduced shear
$g(\bmath{x})$ is defined as
\begin{equation}
g(\bmath{x})=\frac{\gamma(\bmath{x})}{1-\kappa(\bmath{x})},
\label{eq:reduced_shear}
\end{equation}
where the convergence $\kappa(\bmath{x})$ is given by
\begin{equation}
\kappa(\bmath{x})=\frac{\Sigma(\bmath{x})}{\Sigma_{\rm crit}}
\label{eq:convergence}
\end{equation} 
and the shear $\gamma(\bmath{x})$ can, in general, be written as a
convolution integral over the convergence $\kappa(\bmath{x})$
(see e.g. \citealt{bridle99}). $\Sigma_{\rm crit}$ is the critical surface mass
density
\begin{equation}
\Sigma_{\rm crit}=\frac{c^2}{4 \pi G}\frac{D_{\rm s}}{D_{\rm l} D_{\rm ls}}.
\label{eq:sigcrit}
\end{equation}
where $D_{\rm s}$, $D_{\rm l}$ and $D_{\rm ls}$ are the
angular-diameter distances between, respectively, the observer and
each galaxy, the observer and the lens, and the lens and each
galaxy. In general, the redshifts of each background galaxy can be
different, but are assumed to be known.  The lensing effect is said to be
weak or strong if $\kappa \ll 1$ or $\kappa \gtrsim 1$ respectively.
Analytic formulae for the convergence and shear fields produced by a
cluster with an NFW profile have been calculated by
\citet{bartelmann96, wright00, meneghetti03} and we make use of these
here to reduce computational costs.

The observed complex ellipticity components of the $N_{\rm gal}$
background galaxies can be ordered into a data vector $\bmath{d}$
with components
\begin{equation}
\bmath{d}_i = \left\{ 
\begin{array}{ll}
\textrm{Re}(\epsilon_{i}) & \mbox{$\left(i \leq N_{\rm gal}\right)$} \\ & \\
\textrm{Im}(\epsilon_{i-N_{\rm gal}}) & \mbox{$\left(N_{\rm gal}+1 \leq i \leq 2N_{\rm gal}\right)$}
\end{array}
\right.
.
\label{eq:gldvect}
\end{equation}
Likewise the corresponding components of the complex reduced shear
$g(\bmath{x}_i)$ at each galaxy position, as predicted by the cluster
model, can be arranged into the predicted data vector $\bmath{d}^{\rm
  P}$, with the arrangement of components matching Eq.~\ref{eq:gldvect}.

The uncertainty on the measured ellipticity components consists of two
contributions. The intrinsic ellipticity components of the background
galaxies (i.e. prior to lensing) may be taken as having been drawn
independently from a Gaussian distribution with mean zero and variance
$\sigma_{\rm int}^2$. Moreover, the effect of errors in the measured
ellipticity components introduced by the galaxy shape estimation
procedure can be modelled as Gaussian with mean zero and variance
$\sigma^2_{\rm obs}$. Assuming the intrinsic and observational
contributions to the uncertainty are independent, one can simply add
the two variances (see e.g. \citealt{hoekstra00, marshall03}). This
leads to a diagonal noise covariance matrix $\mathbf{C}$ on the
ellipticity components, such that the elements corresponding to the
$i$th galaxy are
\begin{equation}
\sigma^2_{i} = \sigma^2_{\rm obs}+\sigma^2_{\rm int}
[1-\max(|g(\bmath{x}_i)|^2,1/|g(\bmath{x}_i)|^2)]^2.
\label{eq:newsigma}
\end{equation} 
The term inside the square brackets is the correction for the galaxies
lying very close to the critical regions of the strong lensing cluster
as suggested by \citet{schneider00} and implemented by
\citet{bradac04}.

As shown by \citet{marshall03}, we can then write
the likelihood function as
\begin{align}
\mathcal{L}(\mathbf{\Theta})=\frac{1}{Z_L} \exp (
-{\textstyle\frac{1}{2}}\chi^2),
\label{eq:gllhood}
\end{align} 
where $\chi^2$ is the usual misfit statistic 
\begin{align}
\chi^2 
&= \left( \bmath{d} - \bmath{d}^{\rm P} \right)^{\rm T}
\mathbf{C}^{-1} 
\left( \bmath{d} - 
\bmath{d}^{\rm P} \right)\\
&= \sum_{i=1}^{N_{\rm gal}} \sum_{j=1}^{2} \frac{\left(\epsilon_{j,i} -
g_{j}(\bmath{x}_i)\right)^2}{\sigma_i^2},
\label{eq:glchisq}
\end{align} 
and the normalisation factor is
\begin{equation}
Z_{L} = (2 \pi)^{2N_{\rm gal}/2} |\mathbf{C}|^{1/2}.
\label{eq:glchisqnorm}
\end{equation} 
Note that is it necessary to include this normalisation factor in the
likelihood, since the covariance matrix $\mathbf{C}$ is not constant,
but depends on the cluster model parameters through the predicted
shear terms in Eq.~\ref{eq:newsigma}.

\subsection{Priors on cluster parameters}\label{sec:priors}

To determine the model completely it only remains to specify the prior
$\pi(\mathbf{\Theta})$ on the cluster parameters $\mathbf{\Theta} =
(x_{\rm c},y_{\rm c},M_{200},c,z)$. Throughout this paper we assume
the prior to be partly separable, such that
\begin{equation}
\pi(\mathbf{\Theta}) = \pi(x_{\rm c})\pi(y_{\rm c})\pi(c)\pi(M_{200},z).
\label{eq:priordef}
\end{equation}
We assume uniform priors on the position parameters $x_{\rm c}$ and
$y_{\rm c}$
over ranges that slightly exceed the patch of sky for which lensing
data is available, so that the model has the flexibility of having
clusters that lie slightly outside the observed region. This is
necessary since such clusters can still have measurable effects on the
observed shear field. We also adopt the uniform prior
$\pi(c)=\mathcal{U}(0,15)$ on the cluster concentration parameter.

\begin{figure}
\begin{center}
\includegraphics[width=0.8\columnwidth]{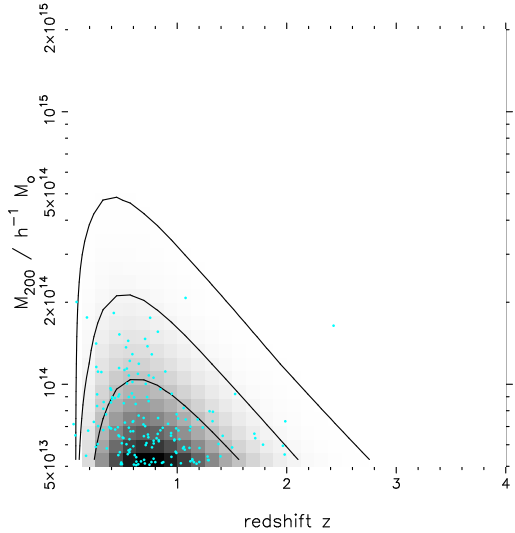}
\caption{The Press-Schechter mass function with $\sigma_8=0.8$,
  together with some samples drawn from it for illustration.  The
  contours enclose 68\%, 95\% and 99\% of the probability.}
\label{fig:press_schechter}
\end{center}
\end{figure}

For the mass $M_{200}$ (which we will denote henceforth by $M$ for
brevity) and redshift $z$ parameters, we adopt a joint prior based on
the Press-Schechter \citep{press74} mass function within some ranges
$M_{\rm min} < M \leq M_{\rm max}$ and $z_{\rm min} < z \leq z_{\rm
  max}$.  The chosen minimum and maximum values of these ranges used
in our analysis of simulated weak-lensing data are discussed below.
Numerical simulations have shown that the Press-Schechter mass
function over-estimates the abundance of high-mass clusters and
under-estimates those of low mass \citep{sheth01}, but overall it
still provides an adequate fit to $N$-body simulations
\citep{jenkins01}. In particular, we assume the Press-Schecter mass
function with $\sigma_8=0.8$, which is plotted in
Figure~\ref{fig:press_schechter}, along with some samples drawn from it
for illustration.
In principle, one could allow $\sigma_8$ to be an additional free
parameter in our model that we attempt to constrain simultaneously
with the parameters describing individual clusters. We have not
pursued this possibility in this first application of our method, but
it will be explored in a future work. It should also be remembered
that our adoption of the Press-Schechter mass function as a prior
still allows for the possibility of detecting clusters with masses and
redshifts that lie outside the ranges favoured by this mass function,
provided the data, through the likelihood function, are sufficiently
conclusive. Finally, we note that is it trivial to replace the assumed
Press-Schechter mass function in our analysis by any other mass
function, if so desired.

\subsection{Quantifying cluster detection}\label{sec:method:bayesian:detection}

As mentioned above, one expects the posterior distribution
$P(\mathbf{\Theta})$ in the parameter space of our single-cluster
model to possess numerous local maxima, some corresponding to real
clusters and some occurring because the pattern of ellipticities in
the background galaxies `conspire' to give the impression that a
cluster might be present. We now discuss how one may calculate the
probability that an identified local peak in the posterior corresponds
to a real cluster (or set of clusters), thereby quantifying our
cluster detection methodology.

This quantification is most naturally performed via a Bayesian model
selection by evaluating the evidence associated with each local
posterior peak for competing models for the data
(see e.g. \citealt{hobson03}). For each peak, it is convenient to consider the
following hypotheses:
\begin{eqnarray*}
H_0 \!\!\!\!\!& = &
\!\!\!\!\! \mbox{`no cluster with $M > M_{\rm lim}$ 
is centred in $S$',}\\
H_1 \!\!\!\!\!& = &
\!\!\!\!\! \mbox{`at least one cluster with $M > M_{\rm lim}$ 
is centred in $S$',}
\end{eqnarray*}
where $S$ is taken to be a region just enclosing (to a good
approximation) the posterior peak in the spatial subspace
$\bmath{x}_c=(x_{\rm c},y_{\rm c})$, and $M_{\rm lim}$ is a lower
limiting mass of interest that we discuss in more detail below. It is
straightforward to see that $H_0$ and $H_1$ are mutually exclusive and
exhaustive hypotheses. The null hypotheses $H_0$ is, however, 
the union of two other mutually exclusive hypothesis, i.e. $H_0=H_2
\cup H_3$, where
\begin{eqnarray*}
H_2 \!\!\!\!\!& = &
\!\!\!\!\! \mbox{`at least one cluster with $0 < M \leq M_{\rm lim}$ 
is centred in $S$',} \\
H_3\!\!\!\!\! & = & 
\!\!\!\!\! \mbox{`no cluster is centred in $S$'.}
\end{eqnarray*}
Note that $H_3$ is equivalent to considering clusters of zero
mass. Moreover, if one choose $M_{\rm lim}=0$, then $H_2$ becomes an
empty hypothesis and $H_0=H_3$.

For each posterior peak, we must calculate the model selection ratio
$R$ given in Eq.~\ref{eq:3.1} between the hypotheses $H_0$ and
$H_1$. Using Bayes' theorem and the fact that $H_0=H_2 \cup H_3$, with
$H_2$ and $H_3$ being mutually exclusive, it is easy to show that $R$
becomes
\begin{equation}
R = \frac{\mathcal{Z}_1}{\mathcal{Z}_2\left[1
+{\displaystyle\frac{\Pr(H_3)}{\Pr(H_2)}}\right]^{-1}
+ \mathcal{Z}_3\left[1+{\displaystyle\frac{\Pr(H_2)}{\Pr(H_3)}}\right]^{-1}}\frac{\Pr(H_1)}{\Pr(H_0)}.
\label{eq:finalrdef}
\end{equation}
Note that for the special case $M_{\rm lim}=0$, for which $P(H_2)=0$ and
$H_0=H_3$, the ratio $R$ correctly reduces to the right-hand side of 
Eq.~\ref{eq:3.1}.

For each hypothesis $H_i$ $(i=1,2,3)$, the evidence is given by
\begin{equation}
\mathcal{Z}_i = \int
\mathcal{L}(\mathbf{\Theta})\pi_i(\mathbf{\Theta})\,d\mathbf{\Theta},
\end{equation}
where 
\begin{equation}
\pi_i(\mathbf{\Theta})=\pi_i(\bmath{x}_{\rm c})\pi_i(c)\pi_i(M,z),
\end{equation}
for $i=1,2,3$ are priors that define the hypotheses. In particular,
the priors on the position of the cluster centre may be taken in all
cases to be uniform: $\pi_i(\bmath{x}_c)=1/|S|$ for all $i$, where
$|S|$ is the area of the region $S$. Similarly, for all hypotheses,
the prior on cluster concentration may be taken
$\pi_i(c)=\mathcal{U}(0,15)$.  Differences between the priors for the
hypotheses do, however, occur in the joint priors $\pi_i(M,z)$ on the
cluster mass and redshift. For hypothesis $H_1$, the prior is the
appropriately normalised Press-Schechter mass function over the
ranges $M_{\rm lim} < M \leq M_{\rm max}$ and $z_{\rm min} < z \leq
z_{\rm max}$, and is zero otherwise.\footnote{Recall that, in any
  case, the Press-Schechter prior is assumed to be zero outside the
  ranges $M_{\rm min} < M \leq M_{\rm max}$ and $z_{\rm min} < z \leq
  z_{\rm max}$, as mentioned in Section~\ref{sec:priors}. Thus, even
  for non-zero $M_{\rm lim}$, if $M_{\rm lim} \leq M_{\rm min}$, then
  $H_2$ becomes an empty hypothesis and $H_0=H_3$.}
 For $H_2$, the prior is the
appropriately normalised Press-Schechter mass function over the
ranges $0 < M \leq M_{\rm lim}$ and $z_{\rm min} < z \leq z_{\rm
  max}$, and is zero otherwise. Finally, for $H_3$, the prior is
$\pi_3(M,z)=\delta(M)\pi_3(z)$, where $\delta(M)$ is the Dirac delta
function centred on $M=0$ and $\pi_3(z)$ can be any normalised
distribution.

Assuming the above priors, the evidence for each hypothesis can be written
\begin{equation}
\mathcal{Z}_i(S) =
\frac{1}{|S|} \int_S \bar{P}_i(\bmath{x}_{\rm c}) \,d^2\bmath{x}_{\rm c},
\label{eq:evidsdef}
\end{equation}
where we have defined the (unnormalised) two-dimensional marginal posterior
\begin{equation}
\bar{P}_i(\bmath{x}_{\rm c})= \int\!\!\int\!\!\int 
\mathcal{L}(\bmath{x}_{\rm c},c,M,z)\,\pi_i(c)
\,\pi_i(M,z)\,dc\,dM\,dz.
\end{equation}
The evidences (Eq.~\ref{eq:evidsdef}) for $i=1,2$ are easily obtained
using the {\sc MultiNest} algorithm, which automatically identifies
local peaks in the posterior and evaluates the `local' evidence
associated with each peak \citep{feroz08,feroz08b}. One minor subtlety
is that, when analysing the weak-lensing survey data, the uniform
prior on the spatial position $\bmath{x}_c$ for clusters extends over
a region $\Omega$ that slightly exceeds the full patch of sky for
which lensing data is available, i.e. $\pi(\bmath{x}_c)=1/|\Omega|$,
as discussed in Section~\ref{sec:priors}). Thus, the local evidence
returned by {\sc MultiNest} as associated with some posterior peak
must me multiplied by $|\Omega|/|S|$ to obtain the corresponding
evidence (Eq.~\ref{eq:evidsdef}).  Finally, the evidence
$\mathcal{Z}_3(S)$ can be calculated directly without the need for any
sampling as
\begin{equation}
\mathcal{Z}_3(S) = \frac{1}{|S|}\int_S \mathcal{L}_0 \,d^2\bmath{x}_c = \mathcal{L}_0,
\end{equation}
since $\mathcal{L}_0\equiv \mathcal{L}(\bmath{x}_c,c,M=0,z)$ is, in fact, independent of
the cluster parameters $c$ and $z$ and the priors are normalised. Note
that $\mathcal{Z}_3$ is, in fact, independent of $S$.

So far we have not addressed the prior ratios $\Pr(H_1)/\Pr(H_0)$ and
$\Pr(H_2)/\Pr(H_3)$ in Eq.~\ref{eq:finalrdef}.  For the sake of
illustration and simplicity, let us assume that the clusters are
randomly distributed in spatial position.  This is not entirely
correct due to clustering of the galaxy clusters on large
scales. Nonetheless, the departure from a random distribution in small
fields is not expected to be significant.  First, consider the ratio
$\Pr(H_1)/\Pr(H_0)$.  If $\mu_S$ is the (in general non-integer)
expected number of clusters with $M > M_{\rm lim}$ centred in a region
of size $|S|$, then the probability of there being $N$ such clusters in
the region is Poisson distributed:
\begin{equation}
\Pr(N|\mu_S) = \frac{e^{-\mu_S} \mu_S^N}{N!}.
\end{equation}
Thus, bearing in mind the above definitions of $H_0$ and $H_1$, we
have
\begin{equation}
\frac{\Pr(H_1)}{\Pr(H_0)} = \exp(\mu_S)-1 \approx \mu_S \mbox{ for
  $\mu_S \ll 1$},
\end{equation}
where $\mu_S$ is given in terms of the mass function
by
\begin{equation}
\mu_{\rm S} = |S|\int_{z_{\rm min}}^{z_{\rm max}} \int_{M_{\rm
    lim}}^{M_{\rm max}} \frac{dn}{dMdz} \,dM\,dz,
\label{eq:num_clusters}
\end{equation}
where $dn/dMdz$ is the distribution of the projected number
density of clusters with masses between $M$ and $dM$ and redshift
between $z$ and $dz$ per unit area.

Let us now consider the ratio $\Pr(H_2)/\Pr(H_3)$
Similarly, let us suppose that $\lambda_S$ is the (in general
non-integer) expected number of clusters with $0<M<M_{\rm lim}$
centered in a region of size $|S|$. Then by the same argument as above
\begin{equation}
\frac{\Pr(H_2)}{\Pr(H_3)} = \exp(\lambda_S)-1 \approx \lambda_S\mbox{ for
  $\lambda_s \ll 1$},
\end{equation}
where $\lambda_S$ is given in terms of the mass function
by
\begin{equation}
\lambda_{\rm S} = |S|\int_{z_{\rm min}}^{z_{\rm max}} \int_{M_{\rm
    min}}^{M_{\rm lim}} \frac{dn}{dMdz} \,dM\,dz.
\label{eq:num_clusters2}
\end{equation}

We are thus able to calculate the model selection ratio $R$ in
Eq.~\ref{eq:finalrdef} for each local posterior peak, which gives us the
relative probability for obtaining a `true' cluster detection ($H_1$)
as opposed to a `false' one ($H_0$). In particular, if the $k$th local
posterior peak has a ratio $R_k$, then the probability that this is a
`true' cluster detection is
\begin{equation} 
p_k = \frac{R_k}{1+R_k}.
\label{eq:prob_TP}
\end{equation}
If we define a `threshold probability' $p_{\rm th}$, such that
detections with $p_k \ge p_{\rm th}$ are identified as candidate
clusters, the expected number of {\it false positives},
$\langle n_{\rm FP}\rangle$ can then be calculated as
\begin{equation}
\langle n_{\rm FP}\rangle = \sum_{k=1, p_k \ge p_{\rm th}}^{K}(1-p_k),
\label{eq:exp_num_FP}
\end{equation}
where $K$ is total number of detected posterior peaks.  The expected
`purity' of the resulting cluster sample, defined as the fraction of
the cluster candidates that are `true', can be
similarly calculated. 

The choice of threshold probability $p_{\rm th}$ depends on the
application. A lower value of $p_{\rm th}$ will obviously result in a
lower purity but higher completeness, especially at the low-mass end
where the lensing signal is particularly weak. In the presence of more
information (e.g. the multi-band photometry) a lower purity but
higher completeness might be preferable and hence, $p_{\rm th}$ should
be set to a lower value in such cases. In the absence of any
additional information for the chosen survey field, $p_{\rm th}$ can
be chosen so that the expected purity is relatively higher. We set
$p_{\rm th}=0.5$ in this work, where we assume no additional
information on the weak lensing simulations under analysis. This
choice of $p_{\rm th}$ ensures that all the detections with a higher
probability of being `true' than being `false' are identified as
candidate clusters. We discuss the impact of $p_{\rm th}$ on the
completeness and purity of the shear selected cluster sample derived
from simulated weak-lensing data in Section~\ref{sec:simpledemo}.

\subsection{Estimation of individual cluster parameters}\label{sec:clustpars}

Once a local posterior peak has been detected and identified as a
cluster candidate using the above method, the values of the parameters
$\mathbf{\Theta}=(\bmath{x}_{\rm c},c,M,z)$ associated with that
cluster are easily calculated using the {\sc MultiNest} algorithm.
The algorithm identifies the samples associated with each posterior
peak and allows one to draw posterior inferences using just these
samples \citep{feroz08,feroz08b}. Thus, for example, one can construct
full one-dimensional marginalised posterior distributions for each
cluster parameter, from which best-fit values and uncertainties are
trivially obtained.

\section{Application to toy weak-lensing simulations}
\label{sec:simpledemo}

\subsection{Simple weak-lensing survey simulation}
\label{sec:simpledemo:simulation}

Before applying our methodology to weak-lensing data derived from full
$N$-body simulations in Section~\ref{sec:whitedemo}, we first
demonstrate its performance in the ideal case, where the clusters have
been simulated using exactly the same model as the one we assume for
the analysis. This is helpful in validating our approach.

\begin{figure*}
\begin{center}
\includegraphics[width=2\columnwidth]{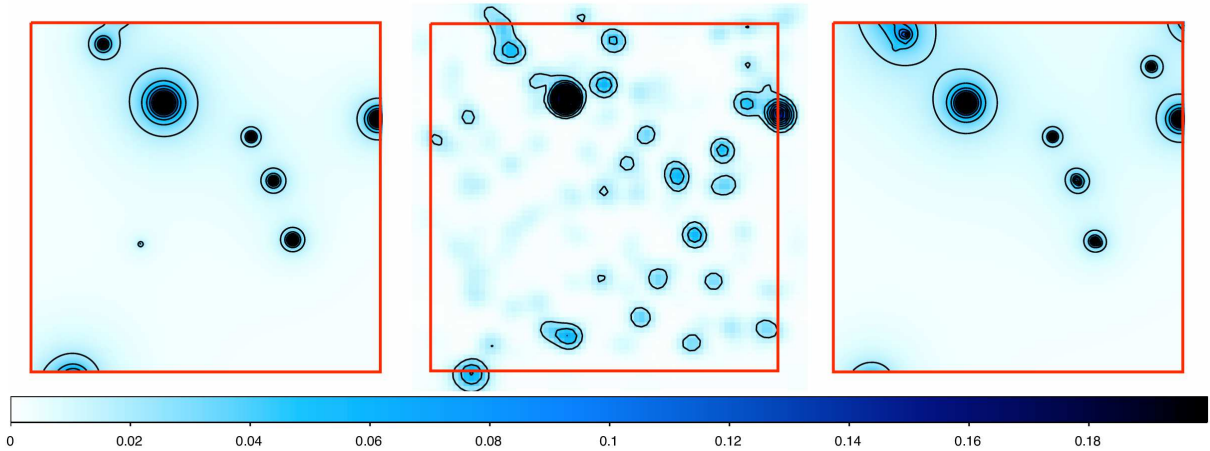}
\caption{Left: true convergence map of the simulated clusters listed
  in Table~\ref{tab:simulation1}. Middle: reconstructed convergence
  map obtained using the LenEnt2 algorithm. Right: inferred
  convergence map obtained using the {\sc MultiNest} algorithm.}
\label{fig:demo1.1}
\end{center}
\end{figure*}

We simulate ten spherically-symmetric clusters, each with an NFW
density profile, distributed uniformly in a $2000 \times 2000$
arcsec$^2$ field, with cluster masses and redshifts drawn from the
Press-Schechter mass function with $M > 5 \times 10^{13} h^{-1}
M_{\sun}$ and $0 \leq z \leq 1$.  We assume a $\Lambda$CDM cosmology
with $\Omega_{\rm m}=0.3$, $\Omega_{\Lambda}=0.7$, $\sigma_8=0.8$ and
Hubble parameter $h=0.7$. The concentration parameter $c$ for each
cluster was drawn from a uniform distribution $\mathcal{U}(0,15)$.
The true cluster parameters are listed in Table~\ref{tab:simulation1}
and the resulting convergence map is shown in Figure~\ref{fig:demo1.1}
(left panel).  

\begin{table}
\begin{center}
\begin{tabular}{rrrrrr}
\hline
 & $x/arcsec$ & $y/arcsec$ & $M_{200} / h^{-1} M_{\sun}$ & $c$ & $z$ \\
\hline
$1$ & $338.9$ & $81.5$ & $7.3 \times 10^{13}$ & $6.2$ & $0.57$ \\
$2$ & $-691.2$ & $-951.4$ & $9.3 \times 10^{13}$ & $6.9$ & $0.21$ \\
$3$ & $-539.1$ & $785.3$ & $5.9 \times 10^{13}$ & $4.0$ & $0.50$ \\
$4$ & $-452.6$ & $963.0$ & $5.0 \times 10^{13}$ & $9.5$ & $0.35$ \\
$5$ & $-345.6$ & $-245.3$ & $5.1 \times 10^{13}$ & $2.4$ & $0.92$ \\
$6$ & $-757.1$ & $-998.7$ & $5.4 \times 10^{13}$ & $2.9$ & $0.47$ \\
$7$ & $877.8$ & $399.9$ & $1.9 \times 10^{14}$ & $9.3$ & $0.47$ \\
$8$ & $-266.5$ & $483.2$ & $3.2 \times 10^{14}$ & $9.4$ & $0.37$ \\
$9$ & $223.6$ & $308.4$ & $5.3 \times 10^{13}$ & $11.1$ & $0.57$ \\
$10$ & $437.9$ & $-222.5$ & $8.6 \times 10^{13}$ & $14.9$ & $0.53$ \\
\hline
\end{tabular}
\caption{True cluster parameters for the simple simulation discussed
  in Section~\ref{sec:simpledemo:simulation}.}
\label{tab:simulation1}
\end{center}
\end{table}

We consider the data only in the middle $1800 \times 1800$ arcsec$^2$
region of the simulation in order to allow the cluster centers to lie
outside the region for which data is available. We down-sample the
resulting convergence and shear maps on a $256 \times 256$ pixel grid
and add Gaussian noise with standard deviation $\sigma_n =
\sigma_{\epsilon}/\sqrt{n_gA}$, where $n_g$ is average number of
galaxies per arcmin$^2$ and $A$ is the pixel area. We assume
$\sigma_{\epsilon} \simeq 0.3$ and $n_g \simeq 100$ gal/arcmin$^2$ as
is approximately found for space-based weak lensing surveys.  We
assume all the background galaxies to lie at redshift $z=1$.

\subsection{Analysis and results}\label{sec:simpledemo:analysis}

We applied our cluster finding methodology, using the {\sc MultiNest} algorithm, to this simple weak lensing
survey simulation. As mentioned in Section~\ref{sec:priors}, we impose uniform priors on cluster positions with
the range of the priors set to be slightly wider than the range for which the data is available. We also
(correctly) impose a uniform prior $\mathcal{U}(0,15)$ on the concentration parameter. For the mass and redshift,
we apply a joint prior coming from the Press-Schechter mass function with $0 \leq z \leq 1$ and $5 \times
10^{13} \leq M / h^{-1} M_{\sun} \leq 5 \times 10^{15}$.

The {\sc MultiNest} algorithm identified $34$ posterior peaks.  For each peak, the probability that it
corresponds to a real cluster was calculated as discussed in Section~\ref{sec:method:bayesian:detection}. $M_{\rm
lim}$ was set to $5 \times 10^{15} h^{-1} M_{\sun}$. Adopting a threshold probability $p_{\rm th} = 0.5$, $10$
candidate clusters were identified. The corresponding inferred convergence map, made from the mean of $100$
convergence maps with cluster parameter values drawn from their respective posterior distributions for the
candidate clusters, is shown in the right-hand panel of Figure~\ref{fig:demo1.1}. For comparison, in the centre
panel we also show the convergence map reconstructed using the LensEnt2 algorithm \citep{marshall03}.

From Figure~\ref{fig:demo1.1} one sees that $7$ of the $10$ candidate clusters correspond to real clusters. We
must however, note that the two overlapping clusters 2 and 6 in Table~\ref{tab:simulation1}, have been confused
to yield a single detected cluster. Of the $10$ candidate clusters, $3$ were actually {\it false positives}, but
we note that $R \sim 1$ for these clusters, corresponding to equal probability of these clusters being either
true or spurious. The $2$ real clusters that were not detected (clusters 4 and 5 in Table~\ref{tab:simulation1})
are the least massive clusters in the simulation and consequently contribute a very small lensing signal. In
fact, cluster 5 was detected by {\sc MultiNest} but had probability ratio $R<1$, corresponding to the
probability of it being `true' of less than 0.5; consequently it was not identified as a candidate cluster with
$p_{\rm th} = 0.5$.

\begin{figure*}
\begin{center}
\scalebox{1.1}{\input{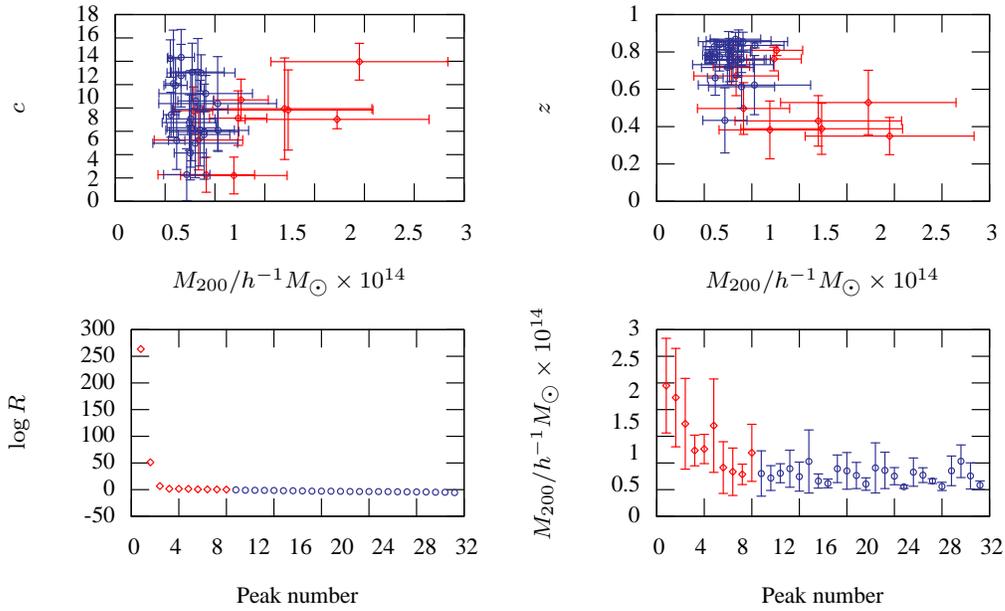}}
\caption{Top panels: inferred parameters values and 68 per cent
  confidence intervals for each of the posterior peaks detected by the
  {\sc MultiNest} algorithm when applied to the simple simulation
  discussed in Section~\ref{sec:simpledemo:simulation}. Bottom panels:
  the $\log R$ value and mass $M$ of detected posterior peak.  In both
  cases, peaks with $R>1$ are plotted in red and the remaining peaks
  in blue.}
\label{fig:demo1.2}
\end{center}
\end{figure*}

\begin{figure*}
\begin{center}
\scalebox{1.1}{\input{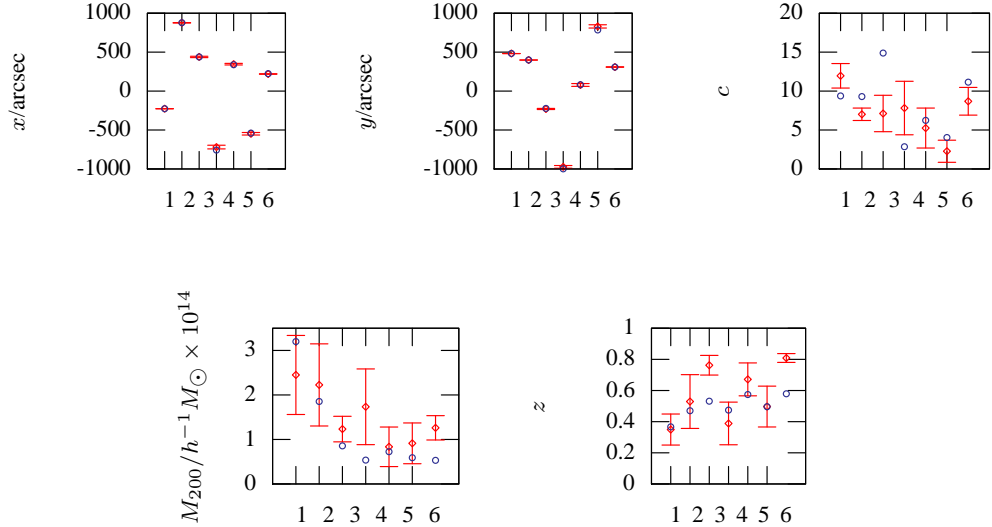}}
\caption{Inferred parameter values with 68 per cent confidence limits (red)
compared to true parameter values (blue) for each candidate cluster.}
\label{fig:demo1.3}
\end{center}
\end{figure*}

The inferred parameter values, with 68 per cent confidence limits, for each of the 32 detected posterior peaks
are shown in Figure~\ref{fig:demo1.2} (top panels). Also shown (bottom panels) is the  $\log R$ value and mass
$M$ of each posterior peak.
Figure~\ref{fig:demo1.3} shows the values of the inferred parameters
(in red) for peaks with $R>1$, and hence identified as cluster
candidates (with $p_{\rm th}=0.5$), together with the true parameter
values (in blue) for the corresponding cluster.

\begin{figure}
\begin{center}
\includegraphics[width=0.75\columnwidth,angle=-90]{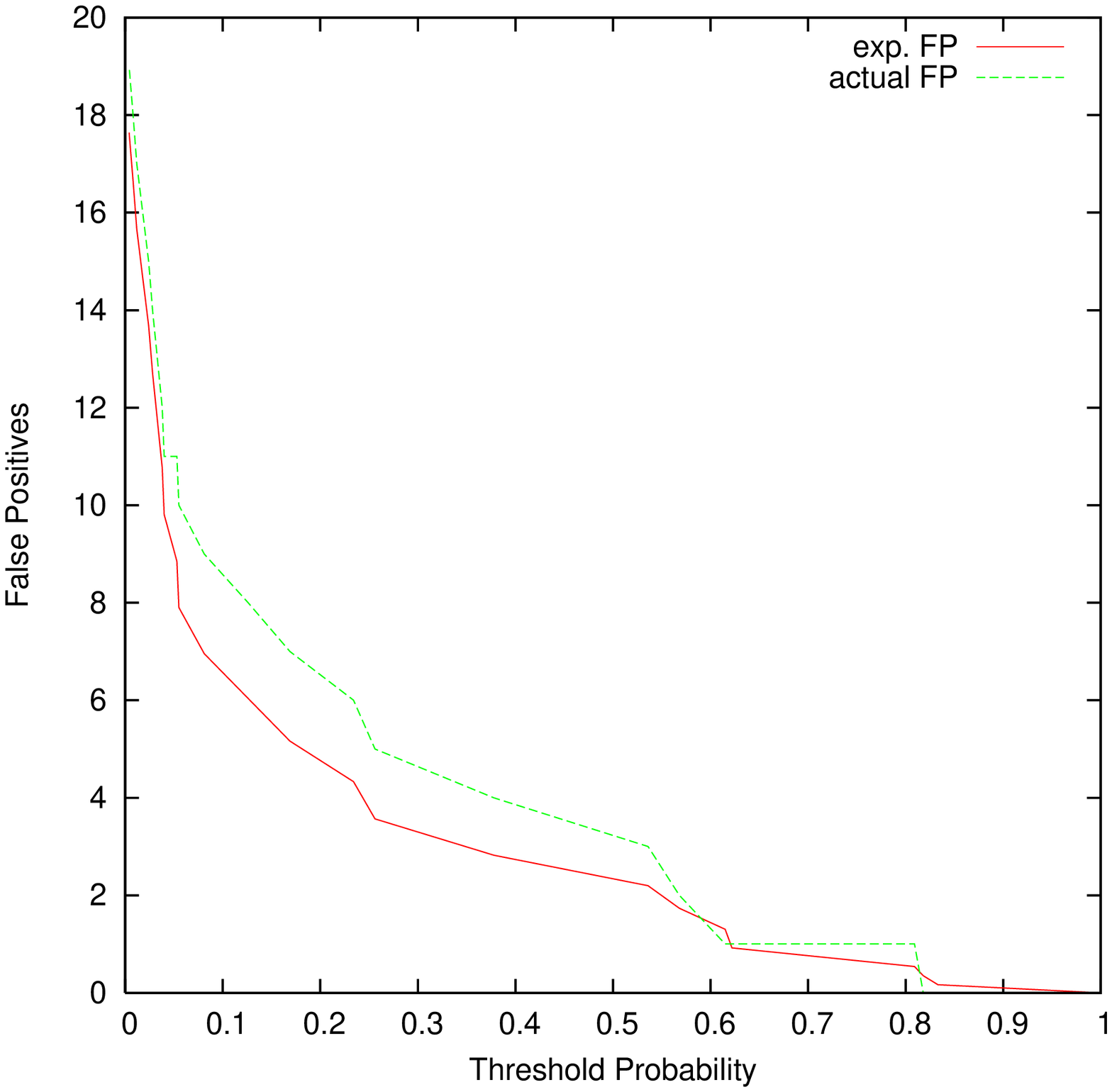}
\caption{The expected and actual number of {\it false positives} as
  function of the `threshold probability' $p_{\rm th}$ (see Section~\ref{sec:method:bayesian:detection}).}
\label{fig:gcat14_FP}
\end{center}
\end{figure}

\begin{figure}
\begin{center}
\includegraphics[width=0.75\columnwidth,angle=-90]{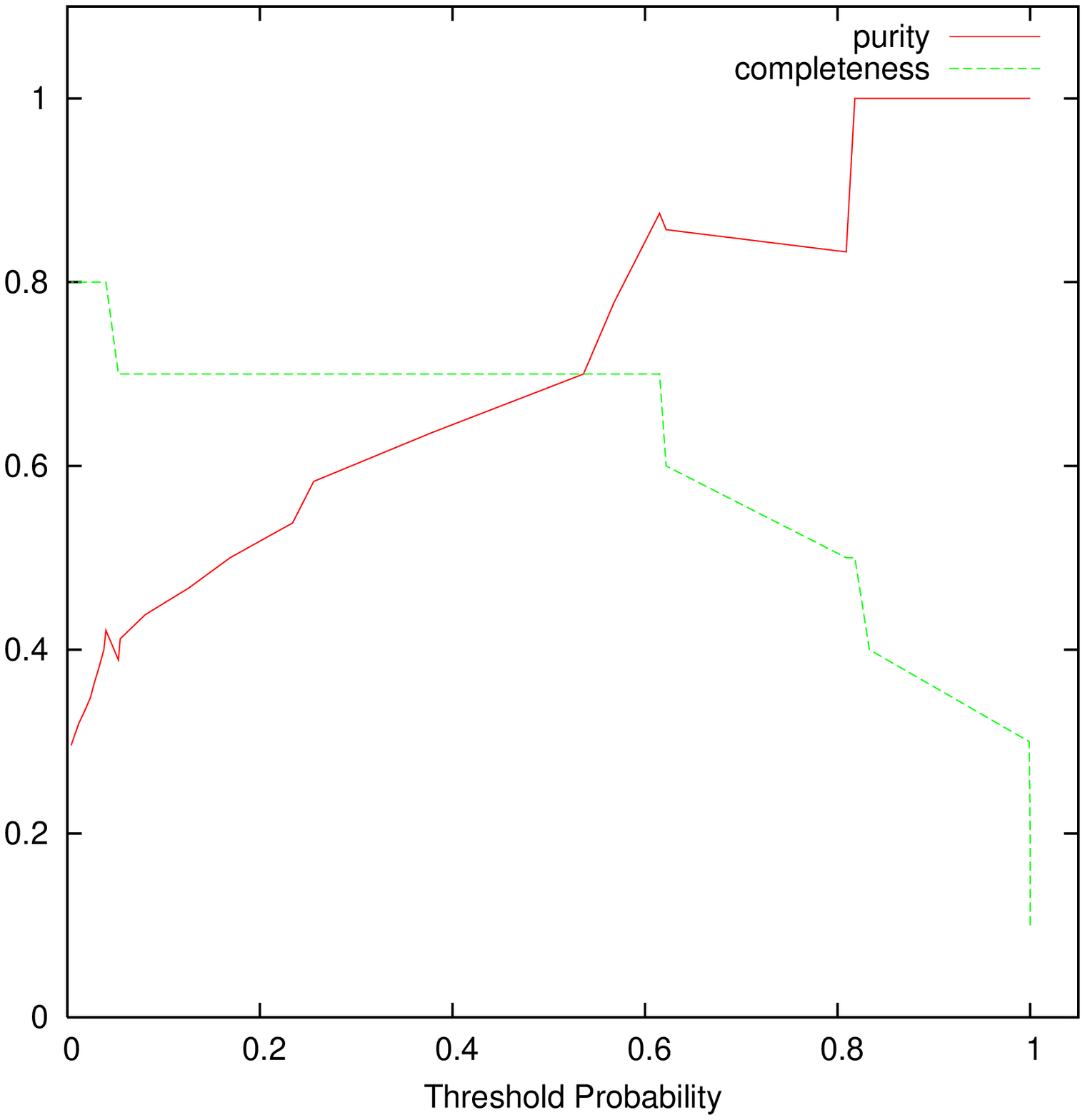}
\caption{Purity and completeness as a function of the `threshold
  probability' for the analysis of the simple simulated weak lensing
  data described in Section~\ref{sec:simpledemo:analysis}.}
\label{fig:gcat14_threshold}
\end{center}
\end{figure}

In Figure~\ref{fig:gcat14_FP}, we also plot the expected
and actual number of {\it false positives} obtained as a function of the
`threshold probability' $p_{\rm th}$. The close agreement between the
two curves indicates that our quantification procedure for cluster
identification is extremely robust.
The corresponding purity and 
completeness as a function of $p_{\rm th}$ are plotted in Figure~\ref{fig:gcat14_threshold}.

\begin{figure}
\begin{center}
\includegraphics[width=0.75\columnwidth,angle=-90]{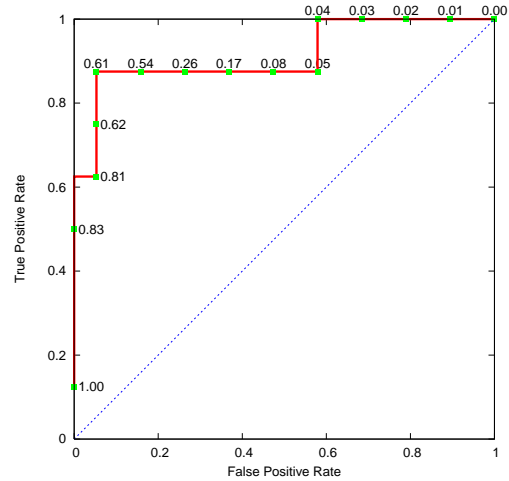}
\caption{Receiver Operating Characteristic (ROC) curve obtained by
  varying the threshold probability $p_{\rm th}$ used in identifying
  candidate clusters for the analysis of the simple simulated weak lensing
  data-set discussed in Section~\ref{sec:simpledemo:analysis}. The
  points are labelled with their corresponding $p_{\rm th}$ value.}
\label{fig:gcat14_ROC}
\end{center}
\end{figure}

Finally, we calculate the Receiver Operating Characteristic (ROC) curve (see e.g. \citealt{fawcett06}) for our
analysis procedure. The ROC curve provides a very reliable way of selecting the optimal algorithm in signal
detection theory. We employ ROC curve here to analyse our cluster candidate identification criterion, based on
the threshold probability $p_{\rm th}$. The ROC curve plots the True Positive Rate (TPR) against the False
Positive Rate (FPR) as a function of the threshold probability. TPR is the ratio of the number of {\it true
positives} for a given $p_{\rm th}$ to the number of {\it true positives} in all the detected clusters (i.e. for
$p_{\rm th}=0$) which for the present analysis is 8 corresponding to 8 clusters that had a strong enough lensing
signal to be identified by our algorithm. Similarly, FPR is the ratio of the number of {\it false positives} for
a given $p_{\rm th}$ to the number of {\it false positives} for $p_{\rm th}=0$. The best possible method would
yield a point in the upper left corner of the ROC space, with coordinates (0,1). A completely random guess would,
on average, yield a point on the diagonal line. The ROC curve traced out as a function of $p_{\rm th}$ for our
cluster detection methodology is plotted in Figure~\ref{fig:gcat14_ROC}. It can bee seen that although, $p_{\rm
th} = 0.6$ gives the optimal cluster candidate identification criteria, $p_{\rm th} = 0.5$ is very nearly as
good. The accuracy of an algorithm can be measured by calculating the area under the ROC curve. An area of $1$
represents a perfect algorithm while an area of $0.5$ represents a worthless algorithm. We notice that the area
under the ROC curve in Figure~\ref{fig:gcat14_ROC} is very close to 1.

%

\section{Application to realistic weak-lensing simulations}
\label{sec:whitedemo}

We now describe the results of our cluster finding algorithm when
applied to simulated weak-lensing survey data derived from numerical
$N$-body simulations.

\subsection{Realistic weak-lensing survey simulation}\label{sec:whitedemo:data}

In this case, the weak-lensing data is simulated using by ray-tracing
through a numerical $N$-body simulation of structure formation
(\citet{white05}), covering a $3 \times 3$ degree field of view. The
cosmological model is taken to be a concordance $\Lambda$CDM model
with parameters $\Omega_{\rm m}=0.3$, $\Omega_{\Lambda}=0.8$,
$\sigma_8=0.9$ and the Hubble parameter $h=0.7$. The simulations
employed $384^3$ equal mass ($10^{10} M_{\sun} h^{-1}$) dark matter
particles in a periodic cubical box of side $200 h^{-1} Mpc$ that was
evolved to $z=0$. The redshift distribution of the background galaxies
was taken to be
\begin{equation}
p(z) \sim z^2 e^{-(z/z_0)^b},
\end{equation}
with $z_0=1.0$ and $b=1.5$.

There are roughly $200$ galaxies per arcmin$^2$ in the simulation,
which represents the deepest space-based surveys. In practice, the
signal-to-noise magnitude threshold gives around $100$ galaxies per
square arcminute and then insisting on knowing the photometric
redshift of each galaxy pushes the number down to $40-70$ per
arcmin$^2$. Since there is no colour-magnitude information for the
galaxies available in these simulations, we pick $65$ galaxies per
square arcminute randomly from the galaxy catalogue. The intrinsic
ellipticity distribution of the galaxies is assumed to be
Gaussian with $\sigma_{int}=0.25$. We add Gaussian observational noise
with standard deviation $\sigma_{obs}=0.20$ to the sheared ellipticity
of each galaxy.

Following \citet{hennawi05}, we define all the halos with $M_{200} >
10^{13.5} h^{-1} M_{\sun}$ as candidates to be identified by
weak-lensing, since the finite number of background galaxies and their
intrinsic ellipticities places a lower limit on the mass of the halo
that can be detected. The halo catalogue for the simulation, produced
using the Friends-of-Friends algorithm (FoF) (\citealt{efstathiou85}),
contains $1368$ halos with $10^{13.5} < M_{200}/h^{-1} M_{\sun} < 7.7
\times 10^{14}$ and $0 < z < 2.7$.

\subsection{Analysis and results}\label{sec:whitedemo:results}

Modelling the entire $3 \times 3$ degree$^2$ field of view represents
a highly computationally expensive task and the {\sc MultiNest}
algorithm would require a prohibitively large number of live points to
be able detect such a great many clusters. We therefore divide the
data into $16$ patches of $0.75 \times 0.75$ degree$^2$ each and use
$4000$ live points to analyse each patch. As before, we expand the
ranges of uniform priors on the position of cluster centers to allow
them to lie a little outside their respective patches, and we use
uniform prior $\mathcal{U}(0,15)$ for the concentration parameter. In
order to determine the effect Press-Schechter mass function prior has
on the detectability of clusters, as well as on the inferred cluster
parameters, we perform the analysis using the joint Press-Schechter
prior on $M_{200}$ and $z$, as well as a logarithmic prior on
$M_{200}$ and uniform prior on $z$. In both cases, we assume the
redshift range $0 < z \leq 4$ and the mass range $10^{13.5} <
M_{200}/(h^{-1} M_{\sun}) \leq 5 \times 10^{15}$.

Before discussing the results from applying {\sc MultiNest} to the
weak-lensing simulation, we note that the `true' cluster catalogue for
$N$-body simulation are, in fact, inferred using a halo finding
algorithm. The FoF algorithm \citet{efstathiou85} is most often
employed for this purpose and works by associating all the particles
within a distance of one another that is some factor $b$ of mean
distance between the particles. The number of halos identified thus
depends very strongly on the linking parameter $b$. Depending on the
resolution of the $N$-body simulation and the value of $b$, FoF may or
may not resolve out structures present within clusters. {\sc
  MultiNest} technique is very sensitive to any structure present
within clusters and will classify these structures as individual
clusters. Therefore, there is always a possibility of {\sc MultiNest}
making an identification of two clusters very close to each other
while the FoF algorithm would identified these as one cluster; this
would result in a lower purity for the {\sc MultiNest} catalogue.

\subsubsection{Completeness and purity}

\begin{figure}
\psfrag{M}[c][c]{$M_{200} h^{-1}M_{\sun}$}
\psfrag{z}[c][c]{$z$}
\begin{center}
\includegraphics[width=1\columnwidth]{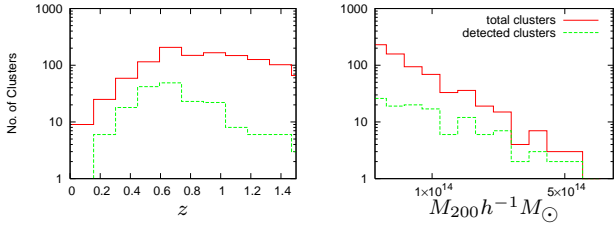}
\caption{The total number of clusters in a given redshift and mass bin
  in the true catalogue (red) against the number of detected clusters
  (green) for the analysis with weak lensing simulation discussed in
  Section~\ref{sec:whitedemo:data} with $p_{\rm th} =
  0.5$. The Press-Schechter mass function was used as a joint prior on
  $M_{200}$ and $z$.}
\label{fig:detections}
\end{center}
\end{figure}

\begin{figure}
\psfrag{M}[c][c]{$M_{200} h^{-1}M_{\sun}$}
\psfrag{z}[c][c]{$z$}
\begin{center}
\includegraphics[width=1\columnwidth]{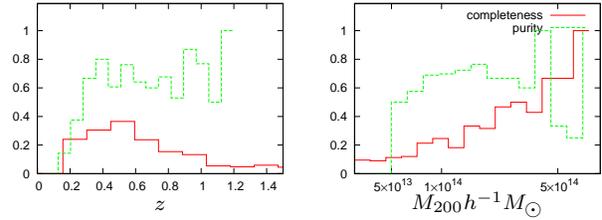}
\caption{Completeness (red) and purity (green) for the analysis with
  weak lensing simulation discussed in Section~\ref{sec:whitedemo:data} with $p_{\rm th} = 0.5$. The
  Press-Schechter mass function was used as a joint prior on
  $M_{200}$ and $z$.}
\label{fig:completeness}
\end{center}
\end{figure}

\begin{figure}
\psfrag{M}[c][c]{$M_{200} h^{-1}M_{\sun}$}
\psfrag{z}[c][c]{$z$}
\begin{center}
\includegraphics[width=1\columnwidth]{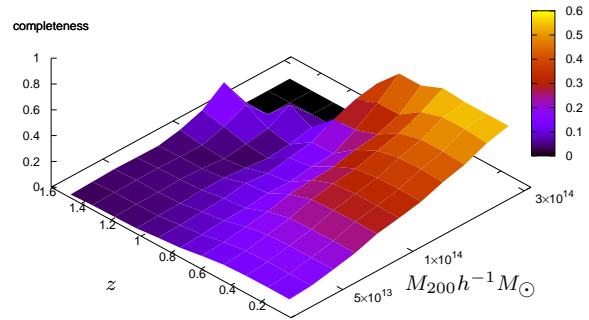}
\caption{Completeness in mass-redshift plane for the analysis of the
  weak-lensing simulation discussed in Section~\ref{sec:whitedemo:data} with $p_{\rm th} = 0.5$. The
  Press-Schechter mass function was used as a joint prior on
  $M_{200}$ and $z$.}
\label{fig:completeness_3d}
\end{center}
\end{figure}

{\sc MultiNest} found around 600 halos out of which 293 and 268 halos
were identified as candidate clusters using $p_{\rm th}=0.5$ with the
Press-Schechter and log-uniform priors respectively. Catalogues
matching was performed for each of these cluster candidates by finding
the closest cluster in the true cluster catalogue for which all the
cluster parameters lie within 4-$\sigma$ of the inferred mean
values. Any cluster candidate not having such a corresponding cluster
in the true catalogue was identified as a \emph{false positive}. Using
this matching scheme, 187 and 175 cluster candidates were identified
as \emph{true positives}, giving a `purity' 64\% and 65\%, for
the Press-Schechter and log-uniform priors respectively. For the
analysis with the Press-Schechter prior, we plot the number of clusters
in the true catalogue and detected clusters as function of true
cluster $M_{200}$ and $z$ in Figure~\ref{fig:detections}. We plot
purity and completeness as a function of cluster $M_{200}$ and $z$ in
Figure~\ref{fig:completeness}. In Figure~\ref{fig:completeness_3d} we
plot the completeness in the mass-redshift plane, for $M_{200} <
3 \times 10^{14} h^{-1} M_{\sun}$ and $z < 1.5$, since there are very
few clusters in the true catalogue outside these ranges and
consequently we suffer from small number statistics.

From Figures~\ref{fig:completeness} and \ref{fig:completeness_3d}, it is
clear that the completeness of our shear-selected cluster sample
approaches unity only for massive clusters with $M_{200} \sim 5 \times
10^{14} h^{-1} M_{\sun}$. \citet{hennawi05} reached a similar
conclusion using the Tomographic Matched Filtering (TMF) scheme.
Unfortunately, a direct comparison with \citet{hennawi05} is not
possible, as they used a different $N$-body simulation. Some other
previous studies of shear-selected cluster samples
\citet{white02,hamana04} have also come to the same conclusion that
they suffer from severe incompleteness except at the high-mass end.
\begin{figure}
\begin{center}
\includegraphics[width=0.75\columnwidth,angle=-90]{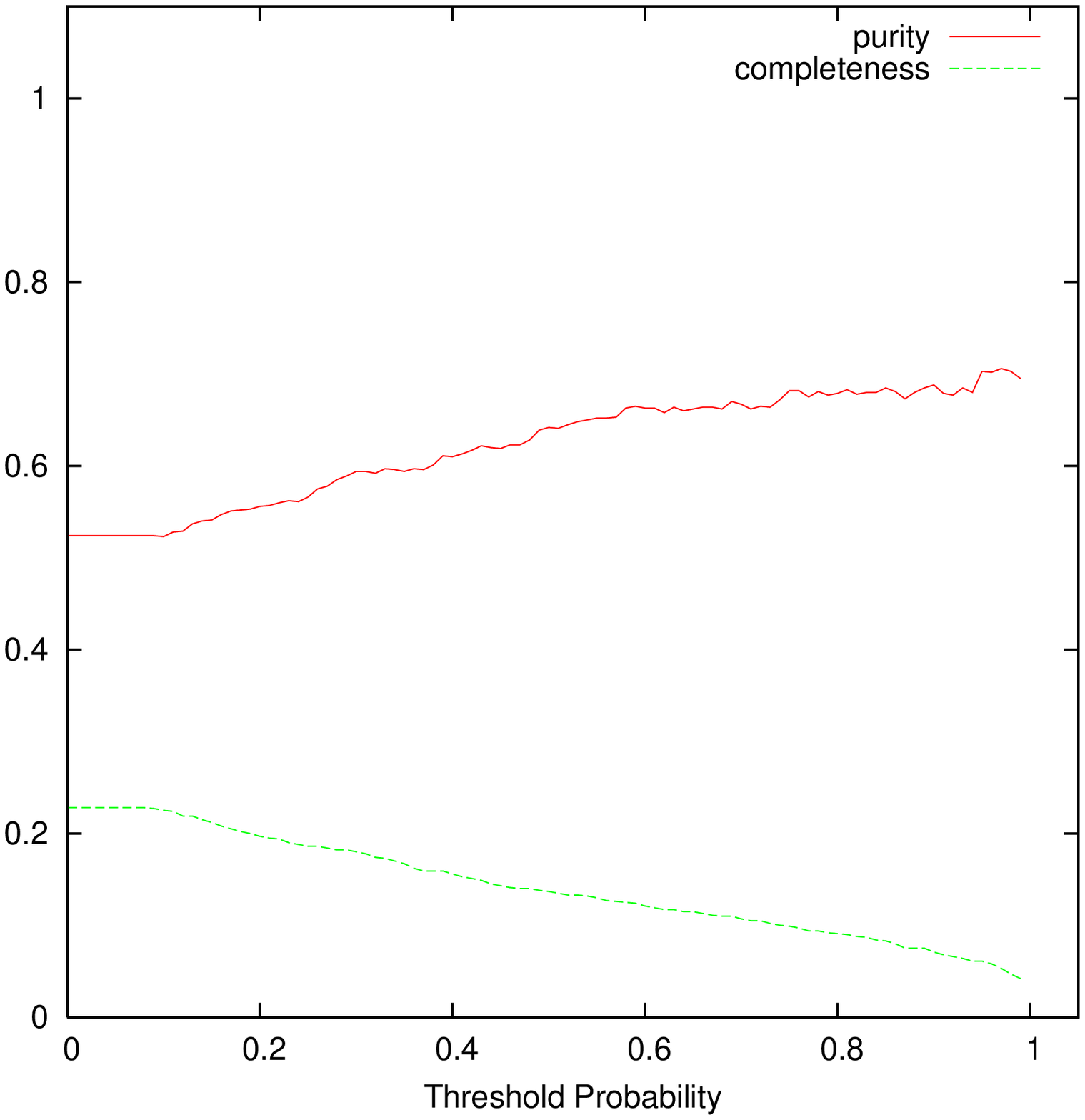}
\caption{Purity and completeness as a function of the `threshold
  probability' for the analysis of the simulated weak-lensing data
  described in Section~\ref{sec:whitedemo:results}. The Press-Schechter
  mass function was used as a joint prior on $M_{200}$ and $z$.}
\label{fig:white_ps_threshold}
\end{center}
\end{figure}

In Figure~\ref{fig:white_ps_threshold} we plot completeness and purity
as a function of threshold probability $p_{\rm th}$ for the analysis
with the Press-Schechter prior. We notice that even for $p_{\rm th}
\sim 1$, purity is around 0.7, while one would expect it to be very
close to unity. This discrepancy occurs because of the presence of
sub-structure in high-mass clusters. As discussed in the previous
section, the {\sc MultiNest} algorithm is very sensitive to any
sub-structure within clusters, and identifies as separate clusters the
halos that the FoF algorithm may or may not identify as belonging to a
single cluster, depending on the impact parameter $b$.

\subsubsection{Accuracy of parameter estimates}

We now discuss the accuracy of recovered parameters of the detected
clusters. The `true' cluster catalogues did not have the concentration
parameter and therefore we do not discuss the accuracy of inferred
concentration of each cluster.

\begin{figure*}
\psfrag{truex}[c][c]{$x$/arcsec}
\psfrag{truey}[c][c]{$y$/arcsec}
\psfrag{truez}[c][c]{$z$}
\psfrag{trueM}[c][c]{$M_{200}/10^{14}h^{-1}M_{\sun}$}
\psfrag{inferredx}[c][c]{$x$/arcsec}
\psfrag{inferredy}[c][c]{$y$/arcsec}
\psfrag{inferredz}[c][c]{$z$}
\psfrag{inferredM}[c][c]{$M_{200}/10^{14}h^{-1}M_{\sun}$}
\begin{center}
\subfigure[]{\includegraphics[width=1.0\columnwidth]{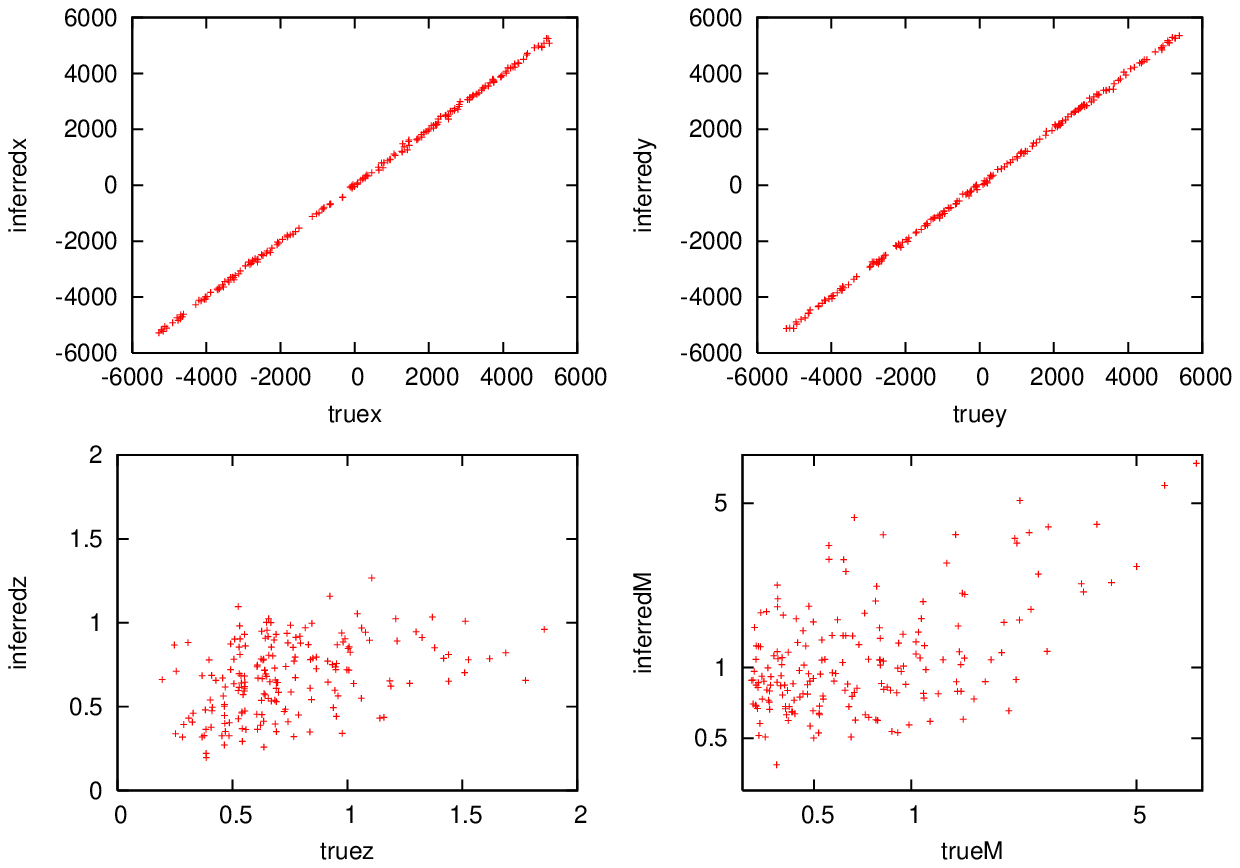}}\hspace{0.3cm}
\subfigure[]{\includegraphics[width=1.0\columnwidth]{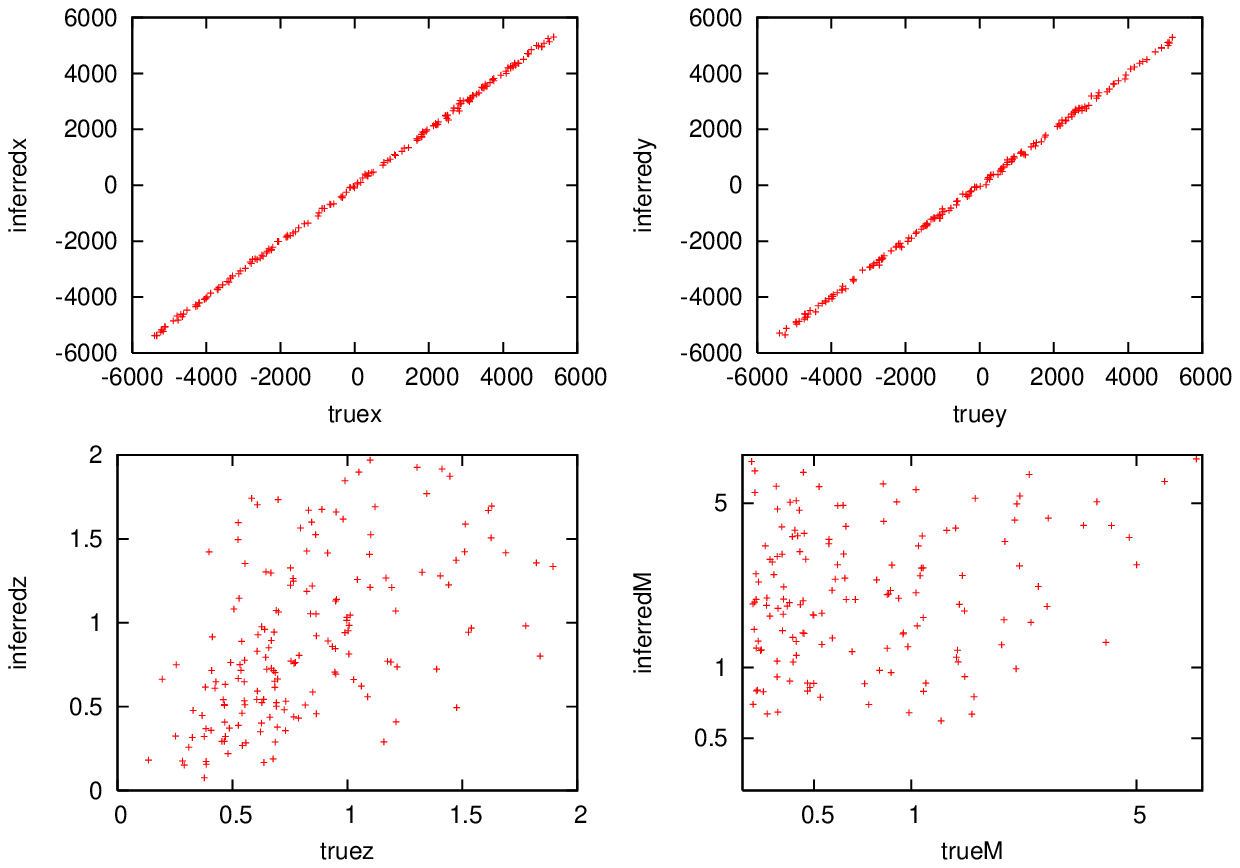}}\hspace{0.3cm}
\caption{Inferred parameter values (on the y-axis) against the `true'
  parameter values (on the x-axis) for each of the detected
  clusters. The analysis was performed with $p_{\rm th}=0.5$ using (a)
  the Press-Schechter mass function prior on $M_{200}$ and $z$ and (b)
  a logarithmic prior on $M_{200}$ and uniform prior on $z$.}
\label{fig:white_par_par}
\end{center}
\end{figure*}
\begin{figure*}
\psfrag{cluster}[c][c]{Cluster No.}
\psfrag{delx}[c][c]{$\Delta x$/arcsec}
\psfrag{dely}[c][c]{$\Delta y$/arcsec}
\psfrag{delz}[c][c]{$\Delta z$}
\psfrag{delM}[c][c]{$\Delta M_{200}$/$10^{14}h^{-1}M_{\sun}$}
\begin{center}
\subfigure[]{\includegraphics[width=1\columnwidth]{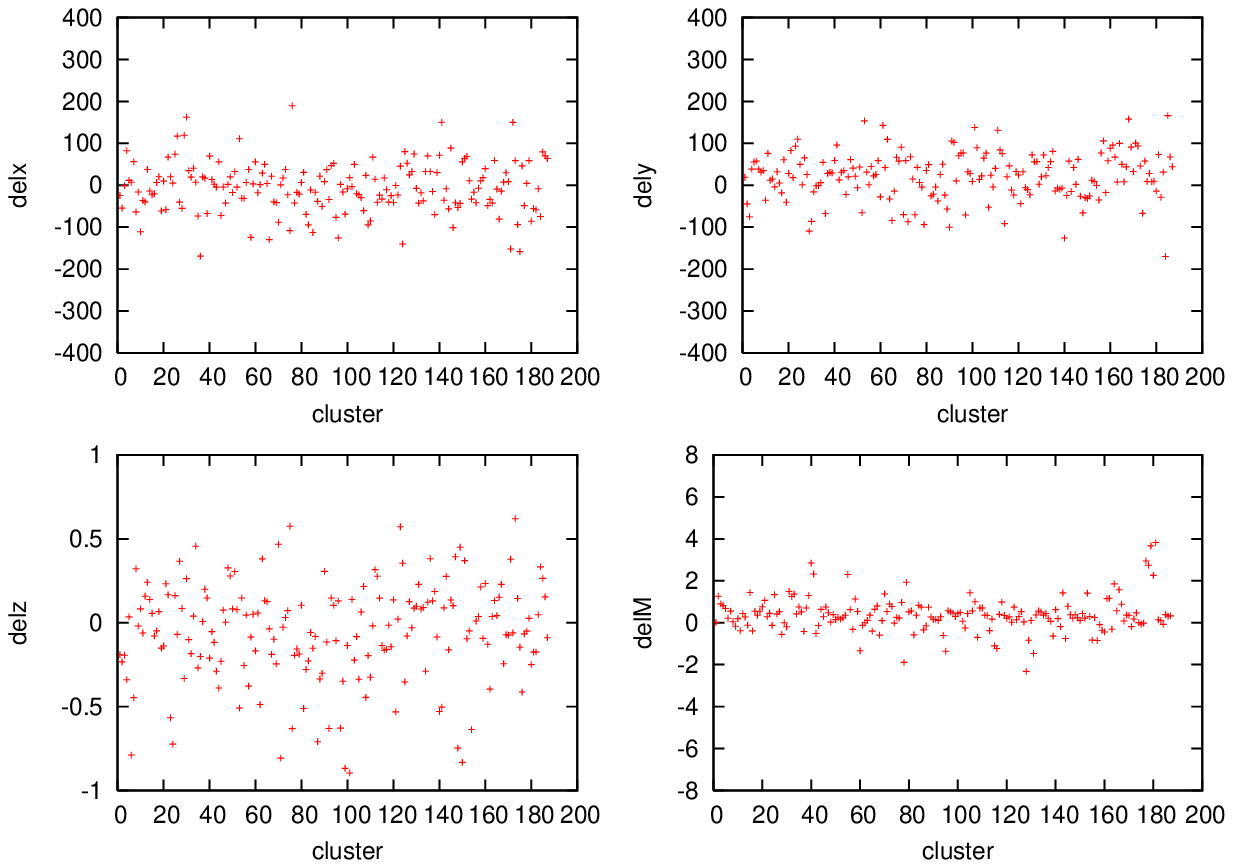}}\hspace{0.3cm}
\subfigure[]{\includegraphics[width=1\columnwidth]{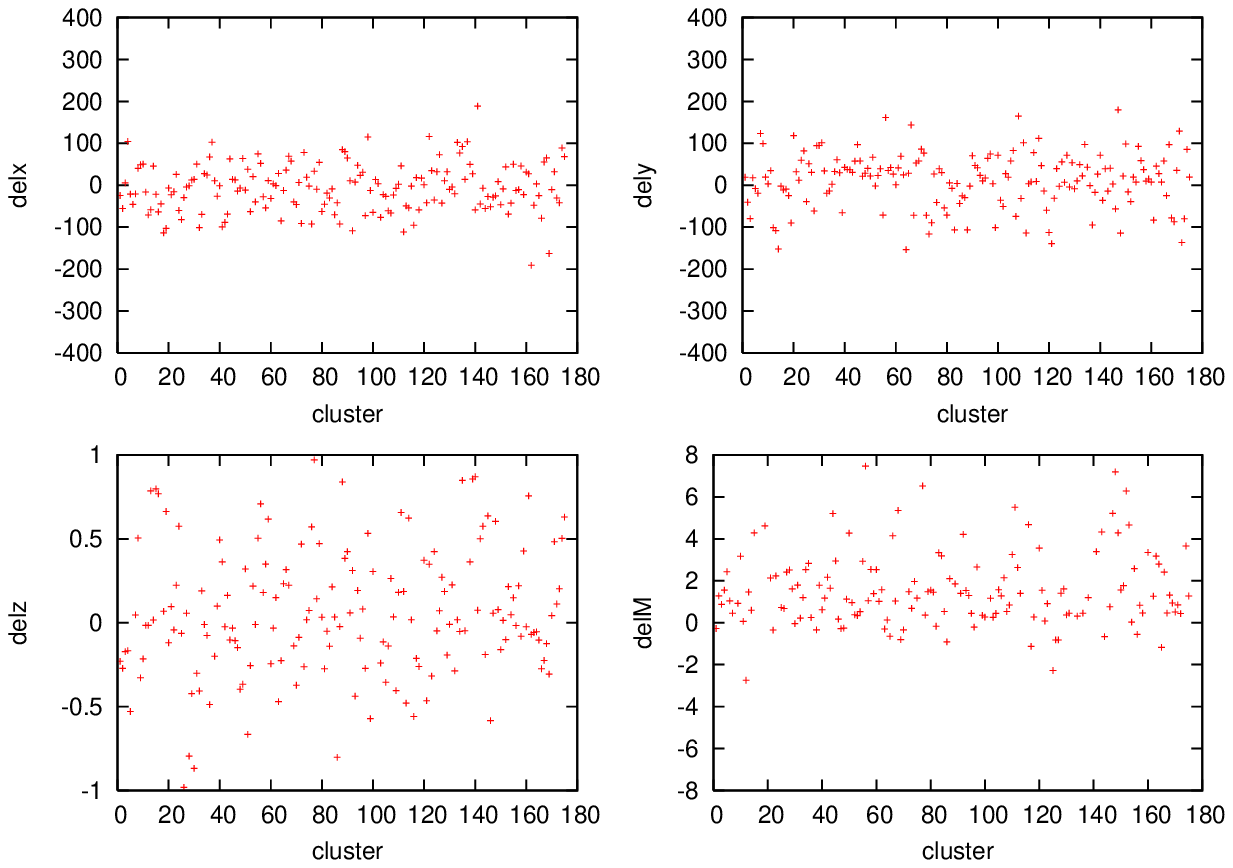}}\hspace{0.3cm}
\caption{Difference between the inferred parameter values and the true
  parameter values from the cluster catalogue, plotted for each
  detected cluster. The analysis was performed with $p_{\rm th}=0.5$ using
  (a) the Press-Schechter mass function prior on $M_{200}$ and $z$ and (b)
  a logarithmic prior on $M_{200}$ and uniform prior on $z$.}
\label{fig:white_par_num}
\end{center}
\end{figure*}
\begin{figure*}
\psfrag{delz}[c][c]{$\Delta z$}
\psfrag{delM}[c][c]{$\Delta M_{200}$/$10^{14}h^{-1}M_{\sun}$}
\psfrag{truez}[c][c]{$z$}
\psfrag{trueM}[c][c]{$M_{200}/10^{14}h^{-1}M_{\sun}$}
\begin{center}
\subfigure[]{\includegraphics[width=1\columnwidth]{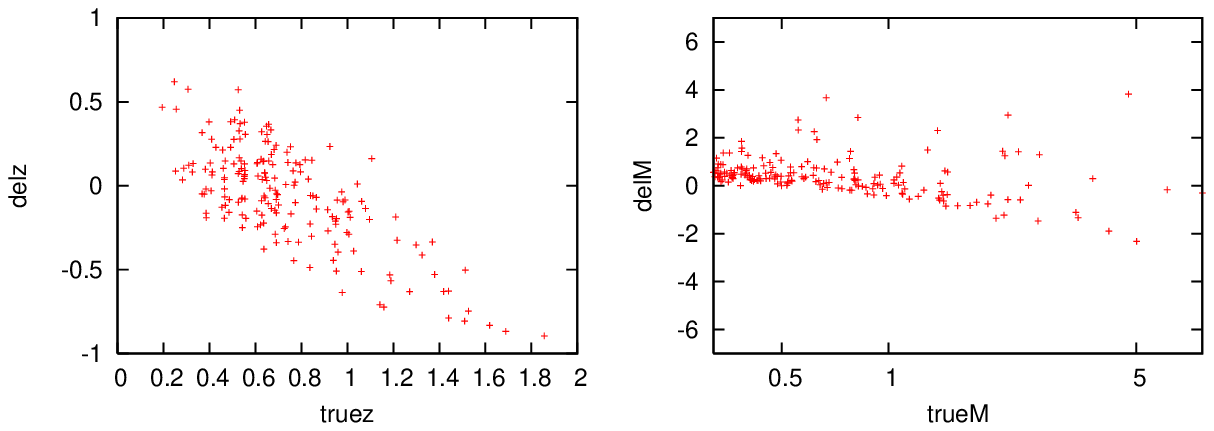}}\hspace{0.3cm}
\subfigure[]{\includegraphics[width=1\columnwidth]{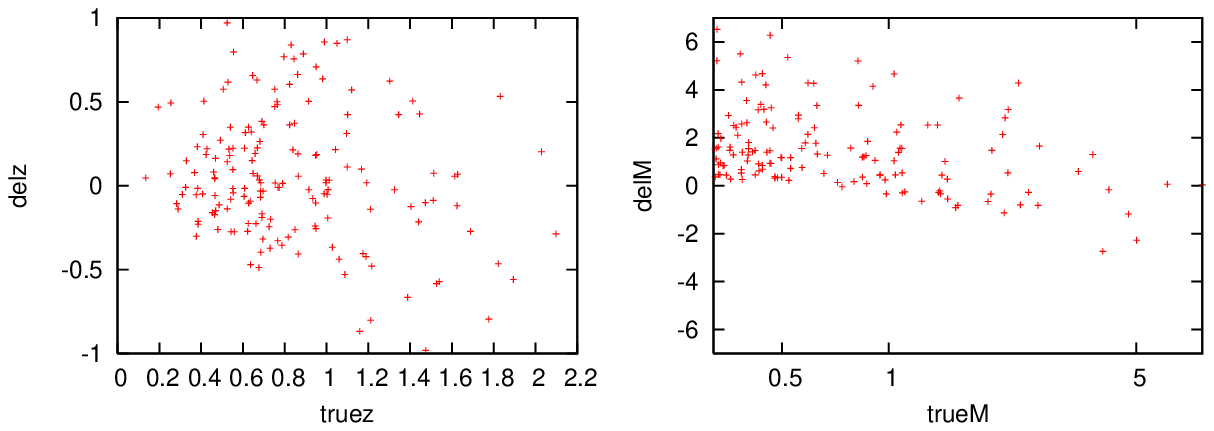}}\hspace{0.3cm}
\caption{Difference between the inferred parameter values and the true
  parameter values from the cluster catalogue, plotted against the
  true parameter values for each of the detected clusters. The
  analysis was performed with $p_{\rm th}=0.5$ using (a) the Press-Schechter
  mass function prior on $M_{200}$ and $z$ and (b) a logarithmic prior
  on $M_{200}$ and uniform prior on $z$.}
\label{fig:white_diff_par_num}
\end{center}
\end{figure*}

For $p_{\rm th}=0.5$, in Figures~\ref{fig:white_par_par}, \ref{fig:white_par_num} and
\ref{fig:white_diff_par_num}, respectively, we plot for each detected cluster: the inferred cluster parameters
against the true cluster parameters; the difference between the inferred parameters and the true parameters; and
the difference between the inferred parameters and the true parameters against the true parameters. Each figure
contains the results for the Press-Schechter and log-uniform priors.

It can be seen from these figures that the cluster positions have been reasonably well estimated, but the
inferred cluster masses have generally been positively biased for both Press-Schechter and logarithmic priors on
$M_{200}$. For logarithmic priors on $M_{200}$ this bias is particularly strong. Applying the Press-Schechter
prior does result in better parameter estimates for cluster masses but nevertheless, the inferred masses of the
low-mass clusters have still been over-estimated. This result agrees well with what is already known from the
$N$-body simulations, the Press-Schechter mass function over-estimates the abundance of high-mass clusters and
under-estimates the abundance of low-mass clusters and consequently, the inferred masses of the low mass clusters
have been over-estimated. This highlights the importance of having physically motivated priors for parameterising
clusters in the weak lensing data-sets. This biasing of cluster masses, particularly at the low-mass end points
to the fact that although the Press-Schechter mass function gives better estimates and reduces the bias in
cluster parameters as compared with a logarithmic prior on $M_{200}$, it still does not fit this particular
$N$-body simulation extremely well. The inferred cluster redshifts have a large scatter for both the
Press-Schechter and logarithmic priors on $M_{200}$. This is because $z$ is correlated with $M_{200}$ and hence
some additional information is required to break this degeneracy and get better estimates for both $M_{200}$ and
$z$.

We further note that we have assumed a spherical NFW model for the cluster mass profiles, while several studies
have shown that the clusters are not necessarily spherical (see e.g. \citealt{shaw06,bett07}).
\citet{corless07,corless08} have recently shown that ignoring the triaxial 3D structure can result in
inaccuracies in cluster parameter estimates. We plan to include the triaxial NFW mass profile in a future work to
assess the importance of including 3D triaxial structure on cluster parameter estimation.

\section{Conclusions}\label{sec:conclusions}

We have introduced a very efficient and robust approach to detecting
galaxy clusters in wide-field weak lensing data-sets.  This approach
allows the parameterisation of each detected cluster. Furthermore,
using Bayesian model selection, one can also calculate the probability
odds for each detected cluster being `true'.  This quantification of
cluster detection allows flexibility in determining the cluster
candidate selection criterion depending on the application. Inspite of
the non-linear nature of the analysis method, we are able to search at
a rate of one-half square degree per hour on a single processor. The
code is fully parallel, making this a viable technique even for the
deepest weak-lensing surveys.

An application of our algorithm to simulated weak-lensing data derived
by an $N$-body simulation showed that the shear-selected cluster
sample suffers from severe incompleteness at the low-mass and
high-redshift ends of the cluster distribution, with the completeness
approaching unity only for massive clusters with $M_{200} \sim 5
\times 10^{14} h^{-1} M_{\sun}$. We also demonstrated the importance
of the priors in estimating the masses and redshifts of low-mass
clusters, since the lensing signal produced by them is particularly
weak. We used the Press-Schechter mass function as a prior on cluster
masses and redshift in this work and found it to be produce inferred
masses of the low-mass clusters that are biassed low.

In a future study, we intend to extend this work by including the
triaxial NFW mass profile to investigate what fraction of clusters
show significant triaxiality in the $\Lambda$CDM model $N$-body
simulation, and also to assess the importance of triaxiality in
cluster parameter estimation.

\section*{Acknowledgements}

This work was carried out largely on the Cambridge High
Performance Computing Cluster Darwin and the authors would like to
thank Dr. Stuart Rankin for computational assistance. FF is supported
by the Cambridge Commonwealth Trust, Isaac Newton and the Pakistan
Higher Education Commission Fellowships.

\bibliographystyle{mn2e}
\bibliography{weak_lensing}

\appendix

\label{lastpage}

\end{document}